\documentclass[reprint, superscriptaddress,amsmath,amssymb, aps, pra,floatfix]{revtex4-2}
\usepackage{nicefrac}
\usepackage{accents}
\usepackage{graphicx} 
\usepackage{dcolumn} 
\usepackage{bm} 
\usepackage[protrusion=true,expansion=true,tracking=true]{microtype}
\usepackage{xcolor}
\usepackage{cancel}
\usepackage{dsfont}
\usepackage{siunitx}
\usepackage[utf8]{inputenc}
\usepackage{hyperref}
\usepackage[capitalize]{cleveref}

\usepackage{soul}

\crefname{section}{Sec.}{Sec.}

\newcommand*\dd{\mathop{}\!\textnormal{\slshape d}}

\newcommand{\imag}{\ensuremath{\text{\textsl{i}}}}

\newcommand{\bra}[1]{\left\langle #1\right|}
\newcommand{\ket}[1]{\left| #1\right\rangle}
\newcommand{\braket}[2]{\ensuremath{\left\langle #1\middle|#2 \right\rangle}}
\newcommand{\expect}[1]{\ensuremath{\langle #1 \rangle}}

\begin{document}

\title{Spin Model for Quantum Annealing with Kerr Parametric Oscillators}
\author{Leo Stenzel}
\email{l.stenzel@parityqc.com}
\affiliation{Parity Quantum Computing Germany GmbH, Schauenburgerstraße 6, 20095 Hamburg, Germany}

\author{Roeland ter Hoeven}
\affiliation{Parity Quantum Computing GmbH, Rennweg 1, Top 314, 6020 Innsbruck, Austria}

\author{Ryoji Miyazaki}
\affiliation{Secure System Platform Research Laboratories, NEC Corporation, Kawasaki, Kanagawa 211-0011, Japan}
\affiliation{NEC-AIST Quantum Technology Cooperative Research Laboratory, Tsukuba, Ibaraki 305-8568, Japan}

\author{Tomohiro Yamaji}
\affiliation{Secure System Platform Research Laboratories, NEC Corporation, Kawasaki, Kanagawa 211-0011, Japan}
\affiliation{NEC-AIST Quantum Technology Cooperative Research Laboratory, Tsukuba, Ibaraki 305-8568, Japan}

\author{Masayuki Shirane}
\affiliation{Secure System Platform Research Laboratories, NEC Corporation, Kawasaki, Kanagawa 211-0011, Japan}
\affiliation{NEC-AIST Quantum Technology Cooperative Research Laboratory, Tsukuba, Ibaraki 305-8568, Japan}

\author{Wolfgang Lechner}
\affiliation{Parity Quantum Computing GmbH, Rennweg 1, Top 314, 6020 Innsbruck, Austria}
\affiliation{Institute for Theoretical Physics, University of Innsbruck, 6020 Innsbruck, Austria}

\date{\today}

\begin{abstract}
Coherent states offer a promising path for near-term quantum computing due to their inherent protection against bit-flip noise.
However, their large photon numbers can be challenging for numerical simulation.
This paper introduces an effective model, representing coherent-state quantum annealing using spin-$1/2$ degrees of freedom.
We demonstrate that this model yields accurate predictions for realistic experimental settings 
and can therefore serve as a practical tool for optimizing future quantum hardware.

\end{abstract}

\maketitle

\section{Introduction}
\label{sec:introduction}
Quantum computing technology has advanced rapidly and has been implemented in several ways. 
One promising platform is superconducting circuits. 
Their high degree of tunability enables us to generate energy structures in which the circuit states are confined to two specific states, 
making them suitable for use as qubits~\cite{Krantz2019}. 
While a representative example of a superconducting qubit relies on the vacuum and single-photon states~\cite{Koch2007}, 
coherent states readily generated in superconducting circuits are also known to function as distinctive qubits due to their robustness against single-photon loss and phase-flip errors~\cite{Cochrane1999,Puri2017a,Guillaud2019}

One prominent application of quantum computations with coherent states is quantum annealing~\cite{Kadowaki1998, Albash2018, Hauke2020} using Kerr parametric oscillators (KPOs) in superconducting circuits~\cite{Goto2016, Puri2017b, Nigg2017, Zhao2018, Onodera2020}. 
In this approach, the KPOs system evolves under parametric driving to a superposition of coherent states, 
whose relative phases correspond to the solution of an embedded Ising problem. 
The efficiency of KPO-based quantum annealing has been numerically demonstrated for small systems~\cite{Goto2016, Puri2017b, Goto2019}. 
This scheme can also be utilized for Boltzmann sampling~\cite{Goto2018} and quantum annealing using excited states~\cite{Zhang2017, Goto2020}. 
In addition, KPO-based quantum annealing is particularly well suited~\cite{Puri2017b} for building the so-called Parity Architecture~\cite{Lechner2015, Ender2023, Drieb-schon2023, Fellner2023, Hoeven2024}, 
which enables problem-specific connectivity, including the all-to-all one, of logical qubits in a planar network of physical qubits. 
The unit of this architecture has recently been implemented in superconducting circuits~\cite{Kawakami2025}.

Numerical simulations play a crucial role in advancing experimental development~\cite{Yamaji2022Feb, Yamaji2022Jun, Yamaji2025, Kawakami2025} of KPO-based quantum annealing. 
They can be utilized to provide theoretical evidence by reproducing experimental results and to optimize experiments by evaluating the impact of each parameter. Simulating the dynamics of KPOs, however, is computationally demanding due to large photon number $n$ in their coherent states. 
For example, a recent experiment for two KPOs was operated at an average photon number of $n=10$, 
requiring up to 30 basis states per KPO when simulated in a Fock basis~\cite{Yamaji2025}. 
The high computational complexity stands in contrast to the fact that this system only has two qubits. 
There should be more resource-efficient numerical approaches that leverage our knowledge of the system’s dynamics, i.e., the time dependence of large-$n$ cat states.

Several numerical methods for handling large-$n$ states have been proposed, but unfortunately, they are unlikely to perform well for simulating KPO-based quantum annealing:
A simple approach shifts the Fock basis states with displacement operators \cite{Chamberland2022}, but, a priori, it
does not work with a time-dependent displacement.
`The dynamic shifted Fock' method \cite{Mercurio2025} can avoid this limitation,
but cannot efficiently handle large-$n$ cat states.
There are also more general approaches that express quantum states in non orthogonal coherent-state bases \cite{Vukics2012,Kramer2011,Schlegel2023}, 
and that compress the local Hilbert-space dimensions in tensor-network states \cite{Koehler2021}.
However, it remains difficult to assess whether these methods can be applied successfully to KPO-based quantum annealing.
Lastly, large Hilbert spaces can be avoided by estimating the evolution of expectation values via a cumulant expansion \cite{Plankensteiner2022}, but this method has been shown to fail for some problems \cite{Kerber2025}.

To overcome these difficulties, this paper derives a simple effective model, describing each KPO using a spin-1/2 degree of freedom.
The model assumes that the states of KPOs are sufficiently close to a basis of coherent states $\{\ket{\pm\alpha}\}$,
and that $\alpha$, including its time dependence, can be estimated from the ideal quantum annealing.
Reducing the system’s degrees of freedom to spin-1/2 significantly simplifies numerical simulations. 
This simplification is also useful for understanding the annealing dynamics.
The present model is simpler than the previously proposed model for KPOs~\cite{Miyazaki2022}, 
yet it provides a more accurate approximation and is better suited for numerical calculations.

This article is organized as follows.
In \cref{sec:method}, we derive the terms of the effective Hamiltonian,
and estimate their time-dependent parameters.
In \cref{sec:res}, we demonstrate that the effective model can predict outcomes for different annealing schedules and realistic experimental parameters.
Finally, we conclude the article in \cref{sec:con}.

Our simulations are implemented in Python using the QuTiP library \cite{Lambert2026},
and the code, and all data shown in this paper are available online \cite{Stenzel2026}.

\section{Method}
\label{sec:method}
This section introduces a spin-$1/2$ Hamiltonian with rescaled coefficients,
which reproduces the dynamics of adiabatic KPO-based quantum annealing.
The spin model has two advantages compared to the exact Fock-space Hamiltonian.
First, the spin model uses a time-independent computational basis, which can be interpreted more easily.
Second, the spin Hamiltonian can be simulated more efficiently on classical hardware:
For the Fock-space Hamiltonian, the number of required basis states per KPO
is determined by the final photon number $\expect{\hat n}\approx \alpha^2$.
For realistic experimental settings, ten or more states may be required per KPO;
simulating a spin-$1/2$ model thus allows us to simulate at least three- or four-times larger systems.  

\subsection{Projection onto Spin States}
The spin model assumes that the state $\ket{\psi(t)}$
of a KPO at time $t$ lies within
the space spanned by coherent states $\ket{\pm \alpha(t)}$.
We explain the estimation of $\alpha$ based on ideal quantum annealing for a specific Hamiltonian in \cref{sec:method:alpha}.

Since $\braket{\alpha}{-\alpha}\neq 0$, we will instead use the cat-state basis for deriving the effective operators,
\begin{align}
    \begin{split}
    \label{eq:simulations:cat+}
    \ket{C_{+}(\alpha)} :=&\; \mathcal{N}_+ \left(\ket{\alpha}+\ket{-\alpha}\right) \\
    = & \; \left(\cosh\alpha^2\right)^{-1/2}\sum_{n=0}^\infty\frac{\alpha^{2n}}{\sqrt{(2n)!}}\ket{2n}_F\,, 
    \end{split}\\
    \begin{split}
    \label{eq:simulations:cat-}
    \ket{C_{-}(\alpha)} :=&\; \mathcal{N}_- \left(\ket{\alpha}-\ket{-\alpha}\right) \\ 
    = & \; (\sinh\alpha^2)^{-1/2} \sum_{n=0}^\infty \frac{\alpha^{2n+1}}{\sqrt{(2n+1)!}} \ket{2n+1}_F\,,
    \end{split}
\end{align}
where $\ket{n}_F$ denotes the Fock-basis state with $n$ particles, and $\mathcal{N}_\pm$ are the respective normalization constants.
To make the distinction between Fock and spin space more clear,
we define the projective transformation as
\begin{equation}
    \hat P(\alpha) = \ket{-}\bra{C_-(\alpha)}\; +\; \ket{+}\bra{C_+(\alpha)}\,,
\end{equation}
where $\{\ket\pm\}$ denote the Pauli-$\hat\sigma^x$ eigenstates of a spin-$1/2$ degree of freedom.
The operator $\hat P(\alpha)$ relates states in a spin-1/2 basis with superpositions of coherent states for a given $\pm \alpha$.
For a Fock state $\ket\psi$, the approximation of the spin model is thus valid if
\begin{equation}
    \label{eq:simulations:assumption}
    \bra{\psi}\hat P^\dagger(\alpha) \hat P(\alpha)\ket{\psi}\approx 1\,.
\end{equation}

In the spin basis, we want to compute the time evolution of
$\ket{\phi(t)} := \hat P\ket{\psi(t)}$, i.e.,
\begin{multline}
    \label{eq:simulations:eom}
    \frac{\dd}{\dd t} \ket\phi =\hat P\frac{\dd}{\dd t}\ket\psi + \left(\frac{\dd}{\dd t}\hat P\right)\ket\psi \approx \\
     \left[\hat P\left(\frac1{\imag\hbar}\hat H_F\right)\hat P^\dagger
        + \left(\frac{\dd}{\dd t}\hat P\right)\hat P^\dagger\right]\ket\phi \equiv \frac1{\imag\hbar} \hat H_S \ket{\phi}\,,
\end{multline}
where 
$\hat H_F$ is the original Hamiltonian, typically represented in the Fock basis, and we use
the approximation corresponding to \cref{eq:simulations:assumption}.
The effective Hamiltonian $\hat H_S$ can be 
further simplified by the fact that the second term is diagonal,
\begin{multline}
    \label{eq:simulations:ppdagger}
    \left(\frac{\dd}{\dd t}\hat P(\alpha)\right)\hat P^\dagger(\alpha)
    = \frac{\dd \alpha}{\dd t}\left(\partial_{\tilde\alpha}\hat P(\tilde\alpha)\right)_\alpha \hat P^\dagger(\alpha) = \\
    \frac{\dd \alpha}{\dd t}\left(\ket+ \partial_{\tilde\alpha}\braket{C_+(\tilde\alpha)}{C_+(\alpha)}\big|_{\tilde\alpha=\alpha} \bra+
        \right.\\ 
        \left.+ \;\ket- \partial_{\tilde\alpha}\braket{C_-(\tilde\alpha)}{C_-(\alpha)}\big|_{\tilde\alpha=\alpha}\bra-\right)\,,
\end{multline}
which can be seen from the definition of the cat states, \cref{eq:simulations:cat+,eq:simulations:cat-}, only
containing either even or odd Fock states.
Furthermore, the diagonal terms in \cref{eq:simulations:ppdagger} also vanish:
This can be seen by computing the derivatives and overlaps explicitly,
or simply by observing that the function
\begin{equation}
    f_s(\tilde\alpha):=\braket{C_s(\tilde\alpha)}{C_s(\alpha)}, \; s\in\{+,-\}, \; \alpha,\tilde\alpha\in \mathbb{R}_{\geq 0}
\end{equation}
has a maximum at $\tilde\alpha = \alpha$, and thus the derivative $\partial_{\tilde\alpha} f_s|_{\alpha}=0$ vanishes.
Thus, the spin-1/2 Hamiltonian can be obtained simply by projection,
and the time-dependence of $\alpha(t)$ does not even matter,
\begin{equation}
    \hat H_S(\alpha) = \hat P(\alpha) \hat H_F\hat P^\dagger(\alpha)\,.
\end{equation}

\subsection{Fock Operators in Spin Basis}
\label{sec:method:operators}
Let us now compute the terms of the KPO Hamiltonian in the spin-1/2 basis explicitly. 
The particle annihilator becomes a combination of two Pauli operators:
\begin{multline}
    \label{eq:simulations:annihilator}
    \hat P(\alpha) \hat a\hat P^\dagger(\alpha)\\
    = \ket-\bra+ \alpha \sqrt{\tanh\alpha^2} + \alpha\sqrt{\coth\alpha^2}\ket +\bra- \\
    =\frac{\alpha}{2}\left[\left(\sqrt{\tanh\alpha^2}-\sqrt{\coth\alpha^2}\right)\imag\hat{\sigma}^y\right.\\\left.+\left(\sqrt{\tanh\alpha^2}+\sqrt{\coth\alpha^2}\right)\hat{\sigma}^z\right]\,.
\end{multline}
The single-photon drive is thus diagonal in the computational basis:
\begin{multline}
    \label{eq:simulations:single_photon}
    \hat P(\alpha) (\hat a+\hat a^\dagger)\hat P^\dagger(\alpha) \\
    =\alpha\left(\sqrt{\tanh\alpha^2}+\sqrt{\coth\alpha^2}\right)\hat\sigma^z\,.
\end{multline}
The particle-number operator corresponds to a Pauli-$\sigma^x$ operator, plus a constant offset:
\begin{multline}
\hat P\hat n\hat P^\dagger = \frac{\alpha^2}{2}\left[\left(\tanh\alpha^2-\coth\alpha^2\right)\hat\sigma^x \right. \\
    \left.+\left(\tanh\alpha^2+\coth\alpha^2\right)\mathds{1}\right]\,.
    \label{eq:simulations:photon_number}
\end{multline}
We note that
\begin{equation}
    \hat P(\alpha) \hat a^2 \hat P^\dagger(\alpha) = \alpha^2\,,
    \label{eq:simulations:pair_annihilation}
\end{equation}
and thus both the two-photon drive,
\begin{equation}
    \hat P(\alpha)\left({\hat a^\dagger}^2 + \hat a^2\right)\hat P^\dagger(\alpha)  = 2\alpha^2\,,
\end{equation}
and the Kerr nonlinearity,
\begin{equation}
    \hat P(\alpha){\hat a^\dagger}^2\hat a^2\hat P^\dagger(\alpha) = \alpha^4\,,
\end{equation}
do not contribute directly to the spin dynamics.

Conversely, \cref{eq:simulations:annihilator,eq:simulations:single_photon,eq:simulations:photon_number,eq:simulations:pair_annihilation}
could be used to express Pauli operators $\hat\sigma$ in terms of Fock-space operators $\hat a$.
This may be useful to engineer a Hamiltonian which realizes specific qubit operations. 
However, the Fock-space system needs to satisfy the assumption \cref{eq:simulations:assumption}, i.e., 
the engineered Hamiltonian has to actually realize coherent states for our approximation to be valid.

\subsection{Estimating \texorpdfstring{$\alpha(t)$}{α(t)}}
\label{sec:method:alpha}
To use the results from \cref{sec:method:operators}, we need to estimate $\alpha(t)$ from the parameters of the Fock-space Hamiltonian.
For concreteness, we consider the 2-KPO Hamiltonian realized in a recent experiment \cite{Yamaji2025}:
\begin{align}
    \hat H_{\mathrm{K}j}(t)=&\,\frac{K_j}{2} {\hat a_j^\dagger}^2 \hat a_j^2 + \Delta(t)\hat n_j + \frac{p_j(t)}{2}\left({\hat a_j^\dagger}^2+\hat a_j^2\right)\,,
    \label{eq:method:h_kpo}\\
    \hat H_{\Omega j}(t) = &\,\Omega_{j}(t)\left(\hat a_j^\dagger + \hat a_j\right)\,,\\
    \hat H_g =&\, g\left(\hat a_0^\dagger\hat a_1 + \hat a_0\hat a_1^\dagger \right)\,,\\
    \hat H(t) =&\,   H_g + \sum_j \hat H_{\Omega j}(t) + \hat H_{\mathrm{K},j}(t)\,,
\end{align}
where the index $j=0/1$ represents the labels of the KPOs, $K_j$ is the Kerr nonlinearity, $p_j$ is the two-photon pump strength, $\Omega_j$ is the one-photon drive strength, and $g$ is the interaction strength between the KPOs.
With the definitions used in Ref.\ \cite{Yamaji2025}, i.e., $K_j<0$, $\Delta(0) < -|g|$, and $p_j(0) = 0$,
the vacuum is the unique maximal-energy state of $\hat H(0)$ \cite{Goto2018}.
For adiabatic annealing, the system should remain in the highest-energy state as the system parameters are varied.
Thus, we expect $\bm\alpha(t)$ to be given by
\begin{equation}
    \bm\alpha(t) \approx \arg\max_{\bm\alpha} \max_\psi \bra{\psi} \hat P(\bm\alpha) \hat H(t) \hat P^\dagger(\bm\alpha)\ket{\psi}\,,
    \label{eq:method:alpha_max}
\end{equation}
where we have to perform a joint optimization for all KPOs simultaneously.
A similar approach, using products of coherent states as trial wave functions, 
was used by \citet{Kanao2021} to estimate the effect of spatially inhomogeneous $\alpha$ on the annealing outcomes.

The maximization in \cref{eq:method:alpha_max} only requires partial diagonalization of the spin Hamiltonian $\hat P\hat H\hat P^\dagger$, 
such that determining $\bm\alpha$ is typically numerically cheaper than actually solving the master equation.

Alternatively, one can assume that the problem Hamiltonian, i.e., the terms $\hat H_{\Omega j}$ and $\hat H_g$, 
are small compared to $\hat H_{\mathrm{K}j}$, such that it suffices to use the bare KPO Hamiltonian \cref{eq:method:h_kpo} in the optimization \cref{eq:method:alpha_max}.
We find this approach yields quite similar results to the following analytical approximation \cite{Goto2018},
especially for larger $p$, i.e., at the end of the annealing process:
\begin{equation}
    \label{eq:simulations:alpha_goto}
    \alpha \approx \sqrt{\frac{\Delta \tanh(p/\Delta)-p}{K}}\,.
\end{equation}

\section{Numerical Results}
\label{sec:res}
This section demonstrates the usefulness of the spin-1/2 model by simulating multi-KPO systems in experimentally realistic regimes,
and compares the results with the ones using the Fock basis.
\Cref{sec:res:2kpo} studies a family of two-KPO systems, closely resembling a recent experiment of some of the authors \cite{Yamaji2025}.
The following \cref{sec:res:4kpo} then shows results for a four-KPO interaction, which could be a building block for quantum optimization using the Parity Architecture \cite{Lechner2015}.

We test the spin model's ability to predict the final state of quantum annealing for different annealing schedules.
While the spin model cannot exactly reproduce the Fock results,
as it cannot represent leakage out of the cat-state basis,
it shows the correct behavior, indicating its usefulness for 
parameter and schedule optimization.

\subsection{2-KPO Simulations}
\label{sec:res:2kpo}
In this section, we simulate a family of two-KPO systems and try to use the 
spin-1/2 model to optimize the schedule of the single-photon drive.
The setup is chosen to mirror recent experiments by \citet{Yamaji2025}:
$K_{0}/2\pi = K_{1}/2\pi = \qty{-12.6}{\mega\hertz}$,
$p_0/2\pi = \qty{148}{\mega\hertz}$,
$p_1/2\pi = \qty{169}{\mega\hertz}$, 
$g/2\pi = \qty{6.9}{\mega\hertz}$,
and $\Delta(0)/2\pi=\qty{-20}{\mega\hertz}$.
The annealing schedule is chosen as in Ref.\ \cite{Yamaji2025}, and stated explicitly in \cref{sec:app:schedule}.
We use $\alpha$ given by \cref{eq:method:alpha_max}.

We also include photon loss and dephasing terms similar to \citet{Yamaji2025},
i.e., simulating the Lindblad equation
\begin{equation}
    \frac{\dd \rho}{\dd t} = -\frac{\imag}{\hbar}\left[\hat H, \rho\right] + \sum_j \left(\frac{\kappa_j}{2}\mathcal{D}[\hat a_j, \rho] + \gamma \mathcal{D}[\hat a^\dagger_j\hat a_j, \rho]\right)\,,
\end{equation}
where the dephasing rate is chosen as $\gamma / 2\pi = \qty{7.7}{\kilo\hertz}$, 
the photon-loss rates as $\kappa_0/2\pi=\qty{1.1}{\mega\hertz}$,
and $\kappa_1/2\pi=\qty{1.3}{\mega\hertz}$,
and
\begin{equation}
    \mathcal{D}[\hat O, \rho] := \hat O\rho\hat O^\dagger - \frac1{2}\left(\hat O^\dagger\hat O \rho + \rho \hat O^\dagger \hat O\right)\,.
\end{equation}
The Lindblad equation for the effective spin model can be obtained using the operators derived in \cref{sec:method:operators},
under the assumption of \cref{eq:simulations:assumption}.

We rescale the strength of the single-photon drive by the maximal $\alpha$, as approximated by \cref{eq:simulations:alpha_goto},
\begin{equation}
    \tilde\Omega_j := \Omega_j \prod_{l\neq j}\sqrt{-K_{l}/p_{l}} \approx \Omega_j \prod_{l\neq j}\alpha_l(T_\mathrm{s})^{-1}\,, 
    \label{eq:rescaled_fields}
\end{equation}
such that $\tilde\Omega$ is comparable with the interaction strength $g$:
For coherent states, the expectation value of the interaction term is $\langle \hat H_g\rangle = 2g\prod_j \alpha_j$, and for the single-photon drive $\langle \hat H_{\Omega j}\rangle = 2 \Omega_j \alpha_l \approx 2\tilde\Omega_j\prod_j\alpha_j$\,.
This approach does not directly translate to KPO systems with multiple interaction terms \cite{Kanao2021}.

\begin{figure}
    \begin{center}
    \includegraphics[width=\columnwidth]{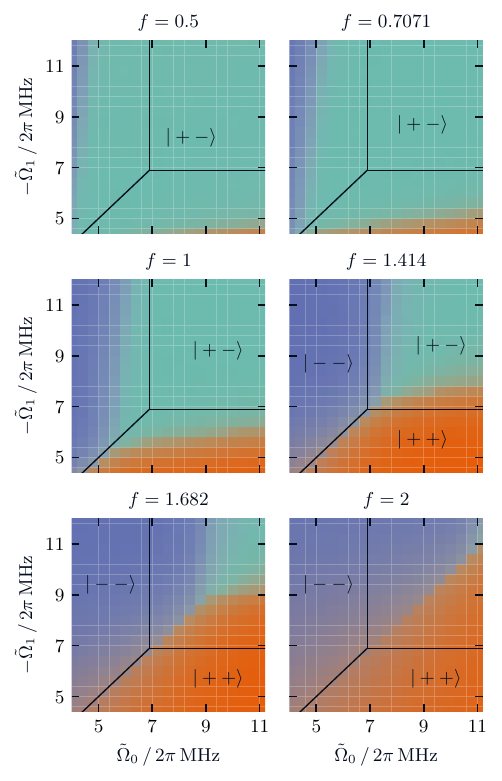}
    \caption{Most likely annealing outcomes as a function of single-photon drive strengths $\tilde{\Omega}_{0,1}$, and single-photon drive schedules, parametrized by $f$.
    Data are computed using the spin model.
    The colors on the heat map represent the most likely final states.
    The black lines indicate the ground-state boundaries of the final Hamiltonian, for all values of $f$.
    The simulations suggest that the outcomes deviate from ideal state boundaries, 
    indicating that our assumption of adiabaticity, used in \cref{eq:method:alpha_max}, can be violated.
    }
    \label{fig:2kpo:diagram}
    \end{center}
\end{figure}

In \cref{fig:2kpo:diagram} we study the case of competition between local fields $\tilde\Omega_j$ and interaction $g$.
We consider $g/2\pi=\qty{6.9}{\mega\hertz}$ which favors parallel `alignment' of the coherent states,
but choose opposite signs for the single-photon drives $\tilde\Omega_0$ and $\tilde\Omega_1$.
In the subplots we observe that the most likely final state
strongly depends on the schedule of the local fields $\tilde\Omega$.
We can understand this by revisiting the motivation behind rescaling the local fields in \cref{eq:rescaled_fields}:
As $\alpha(t)$ changes during the annealing process, the interaction term scales as $\alpha_0(t)\alpha_1(t)$, while the local-field term is proportional to $\langle \hat{H}_{\Omega_j(t)} \rangle \propto t^f \alpha_j(t)$.
For a strictly adiabatic annealing process, the details of the schedule should not matter.
However, we saw in \cref{sec:method:operators} that all terms in the Hamiltonian
become diagonal for large $\alpha$, and thus the system can no longer respond to changes of the local fields.

The spin model is based on two assumptions: 
The instantaneous ground state lies within the coherent-state basis $\ket{\pm\alpha(t)}$,
and $\alpha(t)$ corresponds to the instantaneous maximal energy state.
Neither assumption is strictly true, here:
Due to dephasing, there is leakage out of the coherent-state basis.
And the final states depend on the annealing schedule, meaning that the system does not always realize the maximal energy state.

The fact that the assumptions are not true means that 
the model's validity cannot be guaranteed in the above setting mirroring the experiment. 
However, if the approximations in introducing the spin model have not distorted the schedule dependency, 
practical validity of the model can be found. 
Comparing the simulation results of the spin model and the Fock model should clarify this point.

\begin{figure*}[t]
    \begin{center}
    \includegraphics[width=\textwidth]{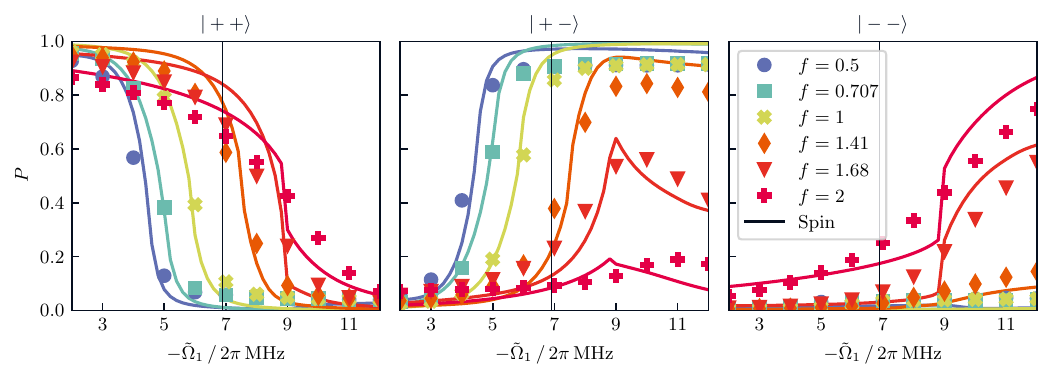}
    \caption{Comparison of Fock and spin simulation for two KPOs.
    Spin simulations correspond to data shown in \cref{fig:2kpo:diagram} for fixed $\tilde{\Omega}_0/2\pi = \qty{9}{\mega\hertz}$.
    Markers indicate Fock simulations, while the solid lines are results of the spin model.
    Probabilities for state $\ket{-+}$ are very close to zero for all methods and parameters, and are thus not shown.
    The black, vertical line indicates the single-photon drive strength $\tilde{\Omega}_1$ where $\ket{++}$ and $\ket{+-}$ are degenerate in energy.
    The spin model tends to overestimate the probability of the most likely spin configuration.
    However, the spin model can quite accurately predict the transition point,
    i.e., the value of $\tilde\Omega_1$ where the most-likely state changes.
    For $f\geq 1.68$ we observe cusps in the spin results, which appear not to be present in the Fock-space simulations. 
    }
    \label{fig:2kpo:spin_fock}
    \end{center}
\end{figure*}

Here, we show that the predictions of the spin model are fairly accurate.
In \cref{fig:2kpo:spin_fock} we compare the spin-model results with simulations in a Fock basis with 27 and 30 basis states for KPO 0 and 1, respectively.
We observe that the spin model tends to overestimate the probability for the 
most likely state.
This is likely a consequence of leakage out of the coherent-state basis due to dephasing $\gamma$, 
which cannot be described accurately in the spin model.
However, the spin-1/2 model accurately predicts the most likely final state,
and, in particular, the transition between different states as function of $\tilde\Omega_1$.
This finding suggests that our spin model is useful for partially optimizing experiments, 
although the valid parameter range is limited, as discussed below.

For larger exponents $f\geq 1.5$ we observe cusps in the spin-model results in \cref{fig:2kpo:spin_fock}, 
which do not appear to exist in the Fock simulations.
We suspect this may be related to our method of estimating $\alpha$, see \cref{eq:method:alpha_max}:
For larger exponents $f$, the local fields $\tilde\Omega$ change more rapidly
at the end of the annealing process;
since $p(t)$, and thus $\alpha(t)$, are larger, the coupling between the $\ket{\pm \alpha}$ states is weak, and the system can no longer follow the maximal energy state.

\subsection{4-KPO Simulations}
\label{sec:res:4kpo}
We want to show the versatility of the spin-1/2 model by simulating a larger system of four KPOs, 
which are interacting as
\begin{equation}
    \hat H_g = g(\hat a^\dagger_0 \hat a^\dagger_1 \hat a_2\hat a_3 + \hat a_0 \hat a_1 \hat a_2^\dagger\hat a_3^\dagger)\,.
\end{equation}
The four-body interaction for KPOs has been studied theoretically~\cite{Puri2017b, Miyazaki2025, Matsuzaki2025}, and has recently been implemented~\cite{Kawakami2025}.
Such a system could, e.g., realize the Parity Architecture \cite{Puri2017b,Lechner2015}.

We use very similar parameters as in \cref{sec:res:2kpo}, and used experimentally by \citet{Yamaji2025}, i.e., a Kerr coefficient of $K/2\pi = \qty{-12.6}{\mega\hertz}$, initial detuning $\Delta(0)/2\pi=\qty{-20}{\mega\hertz}$ ramped linearly to $\Delta(t \geq t_s) = 0$, and a dephasing rate of $\gamma / 2\pi = \qty{7.7}{\kilo\hertz}$. 
We also use comparable photon loss rates $\bm{\kappa} / 2\pi = (0.9, 1.1, 1.3, 1.5)\,\unit{\mega\hertz}$,
but a slightly faster annealing schedule using times $T_\mathrm{s}=\qty{0.5}{\micro\second}$, $T_\mathrm{sp}=\qty{0.1}{\micro\second}$, $T_\mathrm{rd} = \qty{0.1}{\micro\second}$, and $T_\mathrm{r} = \qty{0.1}{\micro\second}$ (see \cref{fig:schedules} for the definition of the parameters).
The shorter total duration is primarily chosen to ease numerical simulation demands. 

Furthermore, to make Fock-space simulations numerically more tractable, 
we reduce the two-photon pump strength to $\bm{p}/2\pi =\left(10, 20, 15, 14\right) \,\unit{\mega\hertz}$, 
and thus reducing the size of the coherent states to $\alpha \in[0.89, 1.26]$.
This reduces the number of required Fock basis states per KPO to about $8$, 
while achieving the same accuracy as in the previous \cref{sec:res:2kpo}.
As a side effect, the reduced $\alpha$ should decrease the influence of dephasing $\gamma$, 
decreasing the leakage out of the $\ket{\pm\alpha}$ basis.
In any case, these parameters would likely not be optimal for future experiments, 
because measuring small-$\alpha$ coherent states would be more difficult, 
and a larger overlap $|\braket{\alpha}{-\alpha}|$ may decrease the fidelity.
We also choose the rescaled single-photon drive strength [see \cref{eq:rescaled_fields}] to be as strong as the four-body interaction,
$\tilde\Omega_0  = \tilde\Omega_2  = -\tilde\Omega_1  =g = 4\pi\,\qty{}{\mega\hertz}$,
and vary the strength of the last KPO's single-photon drive $\tilde{\Omega}_3$.

\begin{figure}[tb]
    \begin{center}
    \includegraphics[width=\columnwidth]{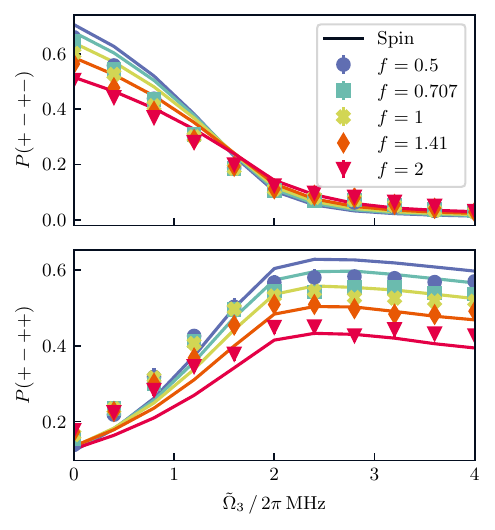}
    \caption{Outcome probabilities for the ideal ground states as a functions of the local single-photon drive on KPO $3$.
    The most likely state changes from $\ket{+-+-}$ to $\ket{+-++}$ for larger $\tilde\Omega_3$.
    Markers correspond to simulations in the Fock space, while the solid lines are the result of simulations in the spin model.
    Compared to \cref{fig:2kpo:spin_fock}, the schedule of the local fields, parametrized by $f$, has a much smaller influence on the observed results.
    }
    \label{fig:4kpo:spin_fock}
    \end{center}
\end{figure}

In \cref{fig:4kpo:spin_fock} we find that the exponent $f$ of the single-photon drive schedule (see \cref{fig:schedules}) has a much smaller impact
on the outcome than what we saw in \cref{sec:res:2kpo}.
We attribute this to the smaller value of $\alpha$:
As the overlap $|\braket{\alpha}{-\alpha}|^2$ is larger, the spin evolution is not `frozen' at the end of the schedule.
And thus, the states at intermediate times during the annealing process do not impact the final outcome as much.
We find the spin model reproduces the correct behavior of the outcome probabilities both as a function of single-photon drive strength $\tilde{\Omega}_3$ and schedule parameter $f$.
Note that the Fock results are obtained by averaging 400 Monte Carlo trajectories \cite{Dalibard1992,Dum1992},
and thus have a statistical error.

\section{Conclusion and Outlook}
\label{sec:con}
In this paper, we propose a simple method for approximating the dynamics of quantum annealers using KPOs.
By projecting onto a time-dependent, two-dimensional basis spanned by coherent states,
we derive an effective Hamiltonian with time-dependent coefficients. 
This allows us to interpret KPO-based quantum annealing in terms of a system of qubits.

The spin model can also be used for more efficient numerical simulations:
For the parameters used in a recent experiment, 
20--30 Fock basis states are required for accurate results.
Thus, the spin model allows us to study systems of more than four times
the number of KPOs with comparable memory and compute requirements.
Note that there is some computational overhead for estimating $\alpha$ for each KPO (see \cref{sec:method:alpha}) but this cost was subleading for all of our simulations.

We show in \cref{sec:res} that the spin model can predict the outcome of different annealing schedules,
and could thus be useful for optimizing future experiments.
There are no free parameters in the effective model,
however the method for estimating $\bm\alpha$ (see \cref{sec:method:alpha})
affects the results.

Further research is required for controlling the spin model's error.
One way could be adding further states to the $\ket{\pm\alpha}$ basis,
similar to other methods using coherent-state bases for numerical simulations 
\cite{Kramer2011, Mercurio2025, Schlegel2023}.
However, such an approach would yield more complicated expressions, 
and require a different implementation.
The error of the spin model can be estimated more simply,
by simulating a subset of KPOs in a Fock basis.
This is particularly useful for single-KPO observables, but it can also serve to compute the leakage out of the cat basis,
giving an insight into the validity of the spin model.

\begin{acknowledgments}
R.M., T.Y., and M.S.\ are grateful to Y.\ Igarashi, Y.\ Kawakami, and T.\ Yamamoto for useful discussions.
L.S.\ thanks P.\ Aumann for useful discussions.
\end{acknowledgments}

\bibliographystyle{apsrev4-2}
\bibliography{bibliography}

@article{Dalibard1992,
  title = {Wave-function approach to dissipative processes in quantum optics},
  author = {Dalibard, Jean and Castin, Yvan and M\o{}lmer, Klaus},
  journal = {Phys. Rev. Lett.},
  volume = {68},
  issue = {5},
  pages = {580--583},
  numpages = {0},
  year = {1992},
  month = {Feb},
  publisher = {American Physical Society},
  doi = {10.1103/PhysRevLett.68.580},
  url = {https://link.aps.org/doi/10.1103/PhysRevLett.68.580}
}

@article{Dum1992,
  title = {Monte Carlo simulation of the atomic master equation for spontaneous emission},
  author = {Dum, R. and Zoller, P. and Ritsch, H.},
  journal = {Phys. Rev. A},
  volume = {45},
  issue = {7},
  pages = {4879--4887},
  numpages = {0},
  year = {1992},
  month = {Apr},
  publisher = {American Physical Society},
  doi = {10.1103/PhysRevA.45.4879},
  url = {https://link.aps.org/doi/10.1103/PhysRevA.45.4879}
}

@article{Cochrane1999,
  title = {Macroscopically distinct quantum-superposition states as a bosonic code for amplitude damping},
  author = {Cochrane, P. T. and Milburn, G. J. and Munro, W. J.},
  journal = {Phys. Rev. A},
  volume = {59},
  issue = {4},
  pages = {2631--2634},
  numpages = {0},
  year = {1999},
  month = {Apr},
  publisher = {American Physical Society},
  doi = {10.1103/PhysRevA.59.2631},
  url = {https://link.aps.org/doi/10.1103/PhysRevA.59.2631}
}

@article{Koch2007,
	title = {Charge-insensitive qubit design derived from the {Cooper} pair box},
	volume = {76},
	url = {https://link.aps.org/doi/10.1103/PhysRevA.76.042319},
	doi = {10.1103/PhysRevA.76.042319},
	number = {4},
	journal = {Phys. Rev. A},
	publisher = {American Physical Society},
	author = {Koch, Jens and Yu, Terri M. and Gambetta, Jay and Houck, A. A. and Schuster, D. I. and Majer, J. and Blais, Alexandre and Devoret, M. H. and Girvin, S. M. and Schoelkopf, R. J.},
	month = oct,
	year = {2007},
	pages = {042319},
}

@mastersthesis{Kramer2011,
    author = {Krämer, Sebastian},
    title = {Simulating open quantum systems with high photon numbers in coherent bases},
    school = {University {I}nnsbruck},
    year = {2011},
    url = {https://ritschgroup.uibk.ac.at/theses/masterthesis_kraemer_sebastian.pdf}
}

@article{Vukics2012,
    title = {C++QEDv2: The multi-array concept and compile-time algorithms in the definition of composite quantum systems},
    journal = {Comput. Phys. Commun.},
    volume = {183},
    number = {6},
    pages = {1381-1396},
    year = {2012},
    issn = {0010-4655},
    doi = {https://doi.org/10.1016/j.cpc.2012.02.004},
    url = {https://www.sciencedirect.com/science/article/pii/S0010465512000562},
    author = {András Vukics},
}

@article{Lechner2015,
author = {Wolfgang Lechner  and Philipp Hauke  and Peter Zoller },
title = {A quantum annealing architecture with all-to-all connectivity from local interactions},
journal = {Sci. Adv.},
volume = {1},
number = {9},
pages = {e1500838},
year = {2015},
doi = {10.1126/sciadv.1500838},
URL = {https://www.science.org/doi/abs/10.1126/sciadv.1500838},
eprint = {https://www.science.org/doi/pdf/10.1126/sciadv.1500838},
}

@article{Goto2016,
	title = {Bifurcation-based adiabatic quantum computation with a nonlinear oscillator network},
	volume = {6},
	issn = {2045-2322},
	url = {https://www.nature.com/articles/srep21686},
	doi = {10.1038/srep21686},
	pages = {21686},
	number = {1},
	journal = {Sci. Rep.},
	author = {Goto, Hayato},
	urldate = {2025-12-04},
	date = {2016-02-22},
    year = {2016},
}

@article{Puri2017a,
	title = {Engineering the quantum states of light in a {K}err-nonlinear resonator by two-photon driving},
	volume = {3},
	issn = {2056-6387},
	url = {https://doi.org/10.1038/s41534-017-0019-1},
	doi = {10.1038/s41534-017-0019-1},
	pages = {18},
	number = {1},
	journal = {npj Quantum Inf.},
	author = {Puri, Shruti and Boutin, Samuel and Blais, Alexandre},
	date = {2017-04-19},
}

@article{Puri2017b,
	title = {Quantum annealing with all-to-all connected nonlinear oscillators},
	volume = {8},
	issn = {2041-1723},
	url = {http://dx.doi.org/10.1038/ncomms15785},
	doi = {10.1038/ncomms15785},
	pages = {15785},
	journal = {Nat. Commun.},
	author = {Puri, Shruti and Andersen, Christian Kraglund and Grimsmo, Arne L and Blais, Alexandre},
	date = {2017-06-08},
}

@article{Zhang2017,
	title = {Preparing quasienergy states on demand: {A} parametric oscillator},
	volume = {95},
	shorttitle = {Preparing quasienergy states on demand},
	url = {https://link.aps.org/doi/10.1103/PhysRevA.95.053841},
	doi = {10.1103/PhysRevA.95.053841},
	number = {5},
	journal = {Phys. Rev. A},
	author = {Zhang, Yaxing and Dykman, M. I.},
	month = may,
	year = {2017},
	pages = {053841}
}

@article{Nigg2017,
author = {Nigg, Simon E. and L{\"{o}}rch, Niels and Tiwari, Rakesh P.},
doi = {10.1126/sciadv.1602273},
issn = {23752548},
journal = {Sci. Adv.},
number = {4},
pages = {e1602273},
title = {{Robust quantum optimizer with full connectivity}},
volume = {3},
year = {2017}
}

@article{Goto2018,
	title = {{Boltzmann sampling from the Ising model using quantum heating of coupled nonlinear oscillators}},
	volume = {8},
	issn = {2045-2322},
	url = {https://www.nature.com/articles/s41598-018-25492-8},
	doi = {10.1038/s41598-018-25492-8},
	pages = {1--9},
	number = {1},
	journal = {Sci. Rep.},
	author = {Goto, Hayato and Lin, Zhirong and Nakamura, Yasunobu},
	urldate = {2022-04-26},
	date = {2018-12-08},
    year = {2018},
}

@article{Zhao2018,
author = {Zhao, Peng and Jin, Zhenchuan and Xu, Peng and Tan, Xinsheng and Yu, Haifeng and Yu, Yang},
doi = {10.1103/PhysRevApplied.10.024019},
issn = {23317019},
journal = {Phys. Rev. Applied},
number = {2},
pages = {024019},
publisher = {American Physical Society},
title = {{Two-Photon Driven Kerr Resonator for Quantum Annealing with Three-Dimensional Circuit QED}},
url = {https://doi.org/10.1103/PhysRevApplied.10.024019},
volume = {10},
year = {2018}
}

@article{Guillaud2019,
  title = {Repetition Cat Qubits for Fault-Tolerant Quantum Computation},
  author = {Guillaud, J\'er\'emie and Mirrahimi, Mazyar},
  journal = {Phys. Rev. X},
  volume = {9},
  issue = {4},
  pages = {041053},
  numpages = {23},
  year = {2019},
  month = {Dec},
  publisher = {American Physical Society},
  doi = {10.1103/PhysRevX.9.041053},
  url = {https://link.aps.org/doi/10.1103/PhysRevX.9.041053}
}

@article{Goto2019,
	title = {Quantum {Computation} {Based} on {Quantum} {Adiabatic} {Bifurcations} of {Kerr}-{Nonlinear} {Parametric} {Oscillators}},
	volume = {88},
	issn = {0031-9015},
	url = {https://journals.jps.jp/doi/full/10.7566/JPSJ.88.061015},
	doi = {10.7566/JPSJ.88.061015},
	number = {6},
	journal = {J. Phys. Soc. Jpn.},
	author = {Goto, Hayato},
	month = mar,
	year = {2019},
	pages = {061015}
}

@article{Goto2020,
	title = {Quantum annealing using vacuum states as effective excited states of driven systems},
	volume = {3},
	copyright = {2020 The Author(s)},
	issn = {2399-3650},
	url = {https://www.nature.com/articles/s42005-020-00502-2},
	doi = {10.1038/s42005-020-00502-2},
	number = {1},
	journal = {Commun. Phys.},
	author = {Goto, Hayato and Kanao, Taro},
	month = dec,
	year = {2020},
	pages = {235}
}

@article{Onodera2020,
	title = {A quantum annealer with fully programmable all-to-all coupling via {Floquet} engineering},
	volume = {6},
	copyright = {2020 The Author(s)},
	issn = {2056-6387},
	url = {https://www.nature.com/articles/s41534-020-0279-z},
	doi = {10.1038/s41534-020-0279-z},
	number = {1},
	journal = {npj Quantum Inf.},
	author = {Onodera, Tatsuhiro and Ng, Edwin and McMahon, Peter L.},
	month = may,
	year = {2020},
	pages = {48}
}

@article{Kanao2021,
  title   = {{High-accuracy Ising machine using Kerr-nonlinear parametric oscillators with local four-body interactions}},
  volume  = {7},
  issn    = {2056-6387},
  url     = {https://doi.org/10.1038/s41534-020-00355-1},
  doi     = {10.1038/s41534-020-00355-1},
  number  = {1},
  journal = {npj Quantum Inf.},
  author  = {Kanao, Taro and Goto, Hayato},
  year    = {2021},
  month   = {Jan},
  pages   = {18}
}

@article{Koehler2021,
  title     = {{Efficient and flexible approach to simulate low-dimensional quantum lattice models with large local Hilbert spaces}},
  volume    = {10},
  issn      = {2542-4653},
  url       = {http://dx.doi.org/10.21468/scipostphys.10.3.058},
  doi       = {10.21468/scipostphys.10.3.058},
  number    = {3},
  journal   = {SciPost Phys.},
  publisher = {Stichting SciPost},
  author    = {Köhler, Thomas and Stolpp, Jan and Paeckel, Sebastian},
  year      = {2021},
  month     = {Mar}
}

@article{Chamberland2022,
  title = {Building a Fault-Tolerant Quantum Computer Using Concatenated Cat Codes},
  author = {Chamberland, Christopher and Noh, Kyungjoo and Arrangoiz-Arriola, Patricio and Campbell, Earl T. and Hann, Connor T. and Iverson, Joseph and Putterman, Harald and Bohdanowicz, Thomas C. and Flammia, Steven T. and Keller, Andrew and Refael, Gil and Preskill, John and Jiang, Liang and Safavi-Naeini, Amir H. and Painter, Oskar and Brand\~ao, Fernando G.S.L.},
  journal = {PRX Quantum},
  volume = {3},
  issue = {1},
  pages = {010329},
  numpages = {117},
  year = {2022},
  month = {Feb},
  publisher = {American Physical Society},
  doi = {10.1103/PRXQuantum.3.010329},
  url = {https://link.aps.org/doi/10.1103/PRXQuantum.3.010329}
}

@article{Plankensteiner2022,
  doi = {10.22331/q-2022-01-04-617},
  url = {https://doi.org/10.22331/q-2022-01-04-617},
  title = {Quantum{C}umulants.jl: {A} {J}ulia framework for generalized mean-field equations in open quantum systems},
  author = {Plankensteiner, David and Hotter, Christoph and Ritsch, Helmut},
  journal = {{Quantum}},
  issn = {2521-327X},
  publisher = {{Verein zur F{\"{o}}rderung des Open Access Publizierens in den Quantenwissenschaften}},
  volume = {6},
  pages = {617},
  month = jan,
  year = {2022}
}

@article{Ender2023,
  doi = {10.22331/q-2023-03-17-950},
  url = {https://doi.org/10.22331/q-2023-03-17-950},
  title = {Parity {Q}uantum {O}ptimization: {C}ompiler},
  author = {Ender, Kilian and ter Hoeven, Roeland and Niehoff, Benjamin E. and Drieb-Sch{\"{o}}n, Maike and Lechner, Wolfgang},
  journal = {{Quantum}},
  issn = {2521-327X},
  publisher = {{Verein zur F{\"{o}}rderung des Open Access Publizierens in den Quantenwissenschaften}},
  volume = {7},
  pages = {950},
  month = mar,
  year = {2023}
}

@misc{Schlegel2023,
      title={{Coherent-State Ladder Time-Dependent Variational Principle for Open Quantum Systems}}, 
      author={David S. Schlegel and Fabrizio Minganti and Vincenzo Savona},
      year={2023},
      eprint={2306.13708},
      archivePrefix={arXiv},
      primaryClass={quant-ph},
      url={https://arxiv.org/abs/2306.13708}, 
}

@article{Drieb-schon2023,
	title = {Parity {Quantum} {Optimization}: {Encoding} {Constraints}},
	volume = {7},
	url = {https://quantum-journal.org/papers/q-2023-03-17-951/},
	doi = {10.22331/q-2023-03-17-951},
	journal = {Quantum},
	author = {Drieb-Sch\"{o}n, Maike and Ender, Kilian and Javanmard, Younes and Lechner, Wolfgang},
	month = mar,
	year = {2023},
	pages = {951},
}

@article{Fellner2023,
	title = {Parity {Quantum} {Optimization}: {Benchmarks}},
	volume = {7},
	url = {https://quantum-journal.org/papers/q-2023-03-17-952/},
	doi = {10.22331/q-2023-03-17-952},
	journal = {Quantum},
	author = {Fellner, Michael and Ender, Kilian and ter Hoeven, Roeland and Lechner, Wolfgang},
	month = mar,
	year = {2023},
	pages = {952},
}

@article{Hoeven2024,
	title = {Constructive plaquette compilation for the parity architecture},
	volume = {9},
	issn = {2058-9565},
	doi = {10.1088/2058-9565/ad5a36},
	number = {3},
	journal = {Quantum Sci. Technol.},
	author = {ter Hoeven, Roeland and Niehoff, Benjamin E. and Kale, Sagar Sudhir and Lechner, Wolfgang},
	month = jul,
	year = {2024},
	pages = {035056},
}

@misc{Kerber2025,
      title={{The Cumulants Expansion Approach: The Good, The Bad and The Ugly}}, 
      author={Johannes Kerber and Helmut Ritsch and Laurin Ostermann},
      year={2025},
      eprint={2511.20115},
      archivePrefix={arXiv},
      primaryClass={quant-ph},
      url={https://arxiv.org/abs/2511.20115}, 
}

@article{Mercurio2025,
  doi = {10.22331/q-2025-09-29-1866},
  url = {https://doi.org/10.22331/q-2025-09-29-1866},
  title = {Quantum{T}oolbox.jl: {A}n efficient {J}ulia framework for simulating open quantum systems},
  author = {Mercurio, Alberto and Huang, Yi-Te and Cai, Li-Xun and Chen, Yueh-Nan and Savona, Vincenzo and Nori, Franco},
  journal = {{Quantum}},
  issn = {2521-327X},
  publisher = {{Verein zur F{\"{o}}rderung des Open Access Publizierens in den Quantenwissenschaften}},
  volume = {9},
  pages = {1866},
  month = sep,
  year = {2025}
}

@misc{Yamaji2025,
      title={{Quantum annealing in capacitively coupled Kerr parametric oscillators using frequency-chirped drives}}, 
      author={T. Yamaji and S. Masuda and Y. Kano and Y. Kawakami and A. Yamaguchi and T. Satoh and A. Morioka and Y. Igarashi and M. Shirane and T. Yamamoto},
      year={2025},
      eprint={2506.23539},
      archivePrefix={arXiv},
      primaryClass={quant-ph},
      url={https://arxiv.org/abs/2506.23539}, 
}

@article{Lambert2026,
    title = {{QuTiP 5: The Quantum Toolbox in Python}},
    journal = {Phys. Rep.},
    volume = {1153},
    pages = {1-62},
    year = {2026},
    issn = {0370-1573},
    doi = {https://doi.org/10.1016/j.physrep.2025.10.001},
    url = {https://www.sciencedirect.com/science/article/pii/S0370157325002704},
    author = {Neill Lambert and Eric Giguère and Paul Menczel and Boxi Li and Patrick Hopf and Gerardo Suárez and Marc Gali and Jake Lishman and Rushiraj Gadhvi and Rochisha Agarwal and Asier Galicia and Nathan Shammah and Paul Nation and J.R. Johansson and Shahnawaz Ahmed and Simon Cross and Alexander Pitchford and Franco Nori},
}

@article{Kadowaki1998,
  title = {Quantum annealing in the transverse Ising model},
  author = {Kadowaki, Tadashi and Nishimori, Hidetoshi},
  journal = {Phys. Rev. E},
  volume = {58},
  issue = {5},
  pages = {5355--5363},
  numpages = {0},
  year = {1998},
  month = {Nov},
  publisher = {American Physical Society},
  doi = {10.1103/PhysRevE.58.5355},
  url = {http://link.aps.org/doi/10.1103/PhysRevE.58.5355}
}

@article{Albash2018,
author = {Albash, Tameem and Lidar, Daniel A.},
doi = {10.1103/RevModPhys.90.015002},
issn = {1539-0756},
journal = {Rev. Mod. Phys.},
number = {1},
pages = {15002},
publisher = {American Physical Society},
title = {{Adiabatic Quantum Computing}},
volume = {90},
year = {2018}
}

@article{Krantz2019,
	title = {A quantum engineer's guide to superconducting qubits},
	volume = {6},
	issn = {1931-9401},
	url = {https://doi.org/10.1063/1.5089550},
	doi = {10.1063/1.5089550},
	number = {2},
	journal = {Appl. Phys. Rev.},
	author = {Krantz, P. and Kjaergaard, M. and Yan, F. and Orlando, T. P. and Gustavsson, S. and Oliver, W. D.},
	month = jun,
	year = {2019},
	pages = {021318},
}

@article{Hauke2020,
title = {Perspectives of quantum annealing: methods and implementation},
volume = {83},
issn = {0034-4885},
shorttitle = {Perspectives of quantum annealing},
url = {https://doi.org/10.1088/1361-6633/ab85b8},
doi = {10.1088/1361-6633/ab85b8},
number = {5},
journal = {Rep. Prog. Phys.},
author = {Hauke, Philipp and Katzgraber, Helmut G. and Lechner, Wolfgang and Nishimori, Hidetoshi and Oliver, William D.},
month = may,
year = {2020},
pages = {054401}
}

@article{Yamaji2022Feb,
	title = {Spectroscopic observation of the crossover from a classical {Duffing} oscillator to a {Kerr} parametric oscillator},
	volume = {105},
	url = {https://link.aps.org/doi/10.1103/PhysRevA.105.023519},
	doi = {10.1103/PhysRevA.105.023519},
	number = {2},
	journal = {Phys. Rev. A},
	author = {Yamaji, T. and Kagami, S. and Yamaguchi, A. and Satoh, T. and Koshino, K. and Goto, H. and Lin, Z. R. and Nakamura, Y. and Yamamoto, T.},
	month = feb,
	year = {2022},
	pages = {023519}
}

@article{Yamaji2022Jun,
	title = {Development of {Quantum} {Annealer} {Using} {Josephson} {Parametric} {Oscillators}},
	volume = {E105-C},
	issn = {1745-1353, 0916-8516},
	url = {https://search.ieice.org/bin/summary.php?id=e105-c_6_283&category=C&year=2022&lang=E&abst=},
	number = {6},
	journal = {IEICE TRANS. ELECTRON.},
	author = {Yamaji, Tomohiro and Shirane, Masayuki and Yamamoto, Tsuyoshi},
	month = jun,
	year = {2022},
	pages = {283--289},
}

@article{Kawakami2025,
    Journal = {arXiv},
	arxivId = {2512.00446}, 
	title = {Four-body interactions in {Kerr} parametric oscillator circuits},
	author = {Kawakami, Yohei and Yamaji, Tomohiro and Yamaguchi, Aiko and Kano, Yuya and Aoki, Takaaki and Taguchi, Aree and Endo, Kiyotaka and Satoh, Tetsuro and Morioka, Ayuka and Igarashi, Yuichi and Shirane, Masayuki and Yamamoto, Tsuyoshi},
	month = nov,
	year = {2025},
	eprint  = {2512.00446}, 
}

@article{Miyazaki2022,
  title = {Effective spin models of Kerr-nonlinear parametric oscillators for quantum annealing},
  author = {Miyazaki, Ryoji},
  journal = {Phys. Rev. A},
  volume = {105},
  issue = {6},
  pages = {062457},
  numpages = {9},
  year = {2022},
  month = {Jun},
  publisher = {American Physical Society},
  doi = {10.1103/PhysRevA.105.062457},
  url = {https://link.aps.org/doi/10.1103/PhysRevA.105.062457}
}

@article{Miyazaki2025,
  title = {Four-body coupler for superconducting qubits based on Josephson parametric oscillators},
  author = {Miyazaki, Ryoji and Yamamoto, Tsuyoshi},
  journal = {Phys. Rev. A},
  volume = {111},
  issue = {6},
  pages = {062612},
  numpages = {16},
  year = {2025},
  month = {Jun},
  publisher = {American Physical Society},
  doi = {10.1103/PhysRevA.111.062612},
  url = {https://link.aps.org/doi/10.1103/PhysRevA.111.062612}
}

@article{Matsuzaki2025,
doi = {10.35848/1347-4065/adcb7a},
url = {https://doi.org/10.35848/1347-4065/adcb7a},
year = {2025},
month = {may},
publisher = {IOP Publishing},
volume = {64},
number = {5},
pages = {052002},
author = {Matsuzaki, Yuichiro and Mori, Yuichiro and Yamaguchi, Aiko and Kawakami, Yohei and Yamamoto, Tsuyoshi},
title = {Theoretical study of the spectroscopic measurements of Kerr nonlinear resonators with four-body interaction},
journal = {Jpn. J. Appl. Phys.},
}

@misc{Stenzel2026,
  author       = {Stenzel, Leo},
  title        = {{Spin-Model Simulations for KPO Annealing}},
  month        = feb,
  year         = 2026,
  journal    = {Zenodo},
  doi          = {10.5281/zenodo.18671233},
  url          = {https://doi.org/10.5281/zenodo.18671233},
}

\appendix

\section{Annealing Schedule}
\label{sec:app:schedule}

\begin{figure}
    \begin{center}
    \includegraphics[width=\columnwidth]{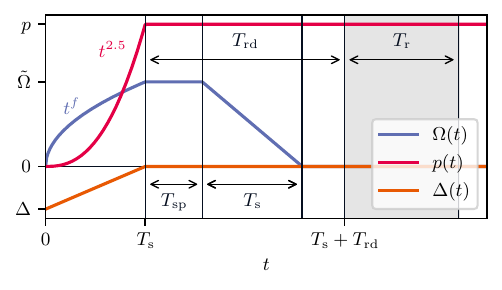}
    \caption{Illustration of the schedules used in our simulations.
    The parameters are taken directly from the recent experiment \cite{Yamaji2025}.
    The state is prepared during $T_\mathrm{s}$, then the 
    single-photon drive $\Omega(t)$ needs to be turned off again,
    before the KPOs are measured during time $T_\mathrm{r}$, indicated
    by the shaded region.
    We use a monomial schedule with exponent $2.5$ for the two-photon drive 
    $P(t)$, as in the experiment \cite{Yamaji2025},
    but vary the exponent $f$ of the local field schedule $\Omega(t)$.
    }
    \label{fig:schedules}
    \end{center}
\end{figure}

The annealing schedule also mimics the experiment~\cite{Yamaji2025}, and is sketched in \cref{fig:schedules}:
One- and two-photon pump strengths are increased in $T_\mathrm{s} = \qty{0.4}{\micro\second}$ to their respective final values.
We scale the two-photon pump as in the experiment as $t^{2.5}$, but consider different monomial schedules for the single-photon drive $\Omega \propto t^f$.
During the same time, the detuning $\Delta(t)$ is linearly increased to zero.

After a hold time of $T_\mathrm{sp} = \qty{0.1}{\micro\second}$, 
the single-photon pump $\Omega(t)$ is again decreased to zero. 
In the experiment, this is required to avoid interference with the readout.
The readout begins at $T_\mathrm{s} + T_\mathrm{rd}$, where $T_\mathrm{rd}=\qty{0.6}{\micro\second}$, and lasts for $T_\mathrm{r} = \qty{0.4}{\micro\second}$.
Numerically, we simulate the readout by time averaging the expectation values over $T_\mathrm{r}$.

\end{document}